\documentclass[12pt]{iopart}

\usepackage{iopams}  
\usepackage{colordvi}
\usepackage{graphicx}% Include figure files
\begin{document}

\title[Polydisperse Colloid-polymer mixtures]{Polydispersity Effects in Colloid-Polymer Mixtures}

\author{S. M. Liddle$^{*}$, T. Narayanan$^{**}$ and W. C. K. Poon$^{*}$}

\address{$^{*}$ SUPA and School of Physics \& Astronomy, The University of Edinburgh, Kings Buildings, Mayfield Road, Edinburgh EH9 3JZ, United Kingdom; $^{**}$ European Synchrotron Radiation Facility, F-38043 Grenoble Cedex, France}
\ead{\mailto{s.liddle@ed.ac.uk}, \mailto{narayan@esrf.fr}, \mailto{w.poon@ed.ac.uk}}
\begin{abstract}
We study experimentally phase separation and transient gelation in a mixture consisting of polydisperse colloids (polydispersity $\approx 6\%$) and non-adsorbing polymers, where the ratio of the average size of the polymer to that of the colloid is $\approx 0.062$. Unlike what has been reported previously for mixtures with somewhat lower colloid polydispersity ($\approx 5\%$), the addition of polymers does {\em not} expand the fluid-solid coexistence region. Instead, we find a region of fluid-solid coexistence which has an approximately constant width but an unexpected re-entrant shape. We detect the presence of a metastable gas-liquid binodal, which gives rise to two-stepped crystallization kinetics that can be rationalized as the effect of fractionation. Finally, we find that the separation into multiple coexisting solid phases at high colloid volume fractions predicted by equilibrium statistical mechanics is kinetically suppressed before the system reaches dynamical arrest.  

\end{abstract}

%Uncomment for PACS numbers title message
%\pacs{00.00, 20.00, 42.10}
% Keywords required only for MST, PB, PMB, PM, JOA, JOB? 
%\vspace{2pc}
%\noindent{\it Keywords}: Article preparation, IOP journals
% Uncomment for Submitted to journal title message
%\submitto{\JPA}
% Comment out if separate title page not required
\maketitle

\section{Introduction}

Well-characterized colloidal suspensions are often used as experimental `models' of phenomena and processes occurring in condensed atomic or molecular systems such as crystallization, various liquid crystalline transitions and glassiness \cite{HNWL02,Poon04}. While this `knowledge transfer' from colloids to the atomic or molecular world works well in many instances, there are also phenomena that are unique to colloids. In particular, real-life synthetic colloids are inevitably polydisperse: the suspended particles are never identical, but have a distribution in properties such as size, shape, charge, etc. Even if many properties are uniform, e.g. in a suspension of uncharged spheres made of a single material, there is an inevitable polydispersity in size. We define this polydispersity, $\sigma$, as the standard deviation of the radius distribution divided by the mean. A bulk sample of polydisperse colloids is, for all practical purposes, an `infinite component system'. In contrast, atomic or molecular materials almost always contain a finite number of components, and the species within each component are (apart from unimportant isotopic variations) identical. 

In colloidal experiments aimed at elucidating generic condensed matter phenomena, polydispersity is usually treated as a problem to be minimized. In such research, the aim is to use particles with as low polydispersity as possible. For example, micron-sized fluorescent silica particles for confocal microscopy can be synthesized with $\sigma \lesssim 2\%$ \cite{Blaaderen92}. A low degree of polydispersity does not seem to introduce qualitatively new phenomena in many instances. Thus, experimentally \cite{Pusey86}, slightly polydisperse hard-sphere colloids ($\sigma \lesssim 5\%$) show the equilibrium phase behaviour predicted by computer simulations for monodisperse hard spheres \cite{Hoover68}, viz., crystallization at volume fractions $\phi \gtrsim 0.5$. Even in this simple case, however, polydispersity can have significant quantitative effects. For example, theory predicts \cite{Warren99,Sollich03} that even at $\sigma$ as low as 3 or 4\%, the melting and freezing volume fractions ($\phi_M$ and $\phi_F$) may be shifted to higher values compared to their monodisperse values of 0.545 and 0.494 respectively. While such predictions are non-trivial to verify because of the difficulty in measuring volume fractions in polydisperse samples, experiments unambiguously demonstrate \cite{vanMegen03} that polydispersity significantly affects kinetics: nucleation was delayed dramatically on increasing $\sigma$ from 4\% to 6.8\%.

Predicting the phase behaviour of polydisperse systems in general, and of polydisperse colloids in particular, poses a significant theoretical challenge. In few-component molecular systems, equating the chemical potentials and pressures of coexisting phases is a practical method for calculations. But such a `brute force' approach becomes increasingly cumbersome as the number of components grows, and entirely new approaches are needed to treat properly the polydisperse, or infinite-component, limit. Significant advances have been made in recent years (reviewed in \cite{Sollich01,Sollich02}), especially on the `reference system' of polydisperse hard spheres. 

In monodisperse hard spheres, increasing $\phi$ of a fluid leads first to fluid-solid (where `solid' means crystal) (FS) coexistence in the interval $0.494 < \phi < 0.545$, and then the equilibrium state is single-phase solid from $\phi = 0.545$ to close packing ($\phi \approx 0.74$) \cite{Hoover68}. We denote this sequence in the self-explanatory notation as `FS $\rightarrow$ S'.  Theory predicts \cite{Sollich03} that a small degree of polydispersity changes the sequence to FS $\rightarrow$ S $\rightarrow$ SS, i.e. at high enough $\phi$, we should find fractionation into coexisting solid phases with different lattice parameters. Above $\sigma \approx 6.5\%$, the single phase solid (S) disappears, and increasing $\phi$ beyond the fluid-solid coexistence region brings the sequence `FS $\rightarrow$ FSS $\rightarrow$ SS $\rightarrow$ SSS'. 

Theoretical predictions have also been made for the effect of polydispersity on the equilibrium phase behaviour of mixtures of hard spheres and non-adsorbing polymers. In such a colloid-polymer mixture, the exclusion of polymer between the surfaces of two nearby particles gives rise to an effective `depletion' attraction whose range and depth are proportional to the polymer size and activity (or, equivalently, chemical potential) respectively \cite{Vrij76}. Since either the polymer or the colloid or both may be polydisperse, a rich variety of effects may be possible in principle. 

The equilibrium phase behaviour of mixtures of monodisperse spherical colloids and polymers is well known (reviewed in \cite{Poon02,Fleer08}) from both theory \cite{Poon92,Fleer07} and experiments \cite{Poon95,Poon08}, and depends on the ratio of the colloid radius ($a$) to the size of the polymer as measured, e.g., by its radius of gyration ($r_g$), $\xi = r_g/a$. At low enough $\xi$, the region of fluid-solid coexistence found at $\phi_F = 0.494 < \phi < \phi_M = 0.545$ for monodisperse hard spheres broadens out. At high enough $\xi$, however, the phase diagram additionally shows regions of gas-liquid and gas-liquid-solid coexistence. The cross over occurs at $\xi = \xi_c$, where theory predicts $\xi_c \approx 0.3$ \cite{Poon92,Fleer08} and experiments find $\xi_c \approx 0.25$ \cite{Poon95}. 

The effect of polymer polydispersity in such mixtures has been studied theoretically \cite{Warren97}. Calculations uncovered no new qualitative effects. In the rest of this paper, we therefore do not consider further the effect of polymer polydispersity.  

Theory suggests that the effect of colloid polydispersity in colloid-polymer mixtures is richer. Fasolo and Sollich \cite{Sollich05} have published schematic representations of the effect of increasing colloid polydispersity, $\sigma$, on colloid-polymer mixture phase diagrams at a range of $\xi$, but have made detailed calculations only for $\xi > \xi_c$. Experimentally, limited data are available for mixtures with $(\xi, \sigma)$= (0.4, 0.18) \cite{Evans98} and (1, 0.54) \cite{Wensink05}. In both cases, significant colloid fractionation was observed in coexisting colloidal gas and liquid phases, with the degree of fractionation obeying a universal law \cite{Evans98}.

While (of necessity) all previous studies of colloid-polymer mixtures in the $\xi < \xi_c$ regime have used polydisperse particles, we are aware of no attempt to date specifically to study the effect of colloid polydispersity on the phase diagram at these smaller size ratios. In this paper, we report the first results of such a study using a hard-sphere-like suspension with $\sigma \gtrsim 6\%$. We find agreement with certain aspects of the phase diagram topology predicted by Fasolo and Sollich \cite{Sollich05}, but that separation into multiple coexisting solid phases is suppressed. These results lead to a slight revision in the interpretation of previously published data on multiple glassy states in a colloid-polymer mixture with a somewhat higher $\sigma$ \cite{Poon02b,Poon04b}. We also detect the presence of a metastable gas-liquid binodal, which gives rise to multi-stepped phase transition kinetics. 

\section{Experimental materials and methods}

\begin{figure}[t]
\begin{center}
\includegraphics[width=0.5\textwidth,clip]{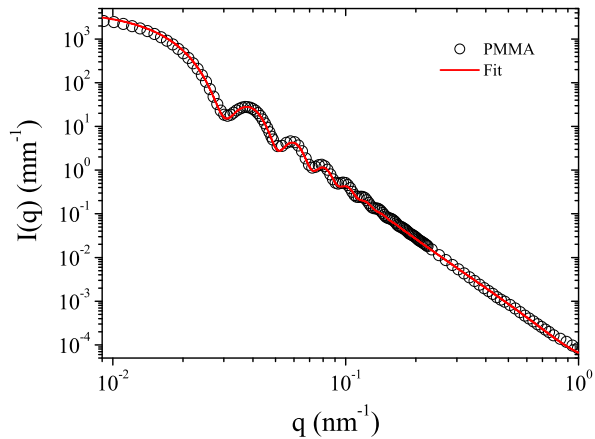}
\end{center}
\caption{Form factor of PMMA colloids (points) from SAXS fitted with a polydisperse core-shell model convoluted with an instrument resolution function of width $0.0026$~nm$^{-1}$
 (continuous line, red online).}
\label{form}
\end{figure}

We used particles of polymethylmethacrylate (PMMA) sterically stabilised by chemically-grafted poly-12-hydroxystearic acid (PHSA) \cite{Antl} synthesized in house and suspended in a mixture of cis- and trans-decalin (density 0.88123~g/cm$^3$). We characterized these particles using small-angle X-ray scattering (SAXS) (reviewed in \cite{Balluff96}). SAXS measurements were carried out at the ID02  beamline at the European Synchrotron Radiation Facility in Grenoble (France). Two sample-to-detector distances of 10~m and 1~m were used to cover a scattering vector ($q$) range of $0.008~{\rm nm}^{-1} \leq q \leq 3~{\rm nm}^{-1}$.  Colloidal suspensions were contained in a flow-through capillary cell of diameter 1.8~mm. Measured two-dimensional scattering patterns were normalized to an absolute intensity scale and then azimuthally averaged to obtain the intensity as a function of $q$, $I(q)$ \cite{Narayanan09}. Fig.~\ref{form} shows the background-subtracted SAXS intensity from a dilute suspension. The data can be adequately described by a polydisperse core-shell model with a Schulz size distribution \cite{Narayanan09}. The best fit parameter values are: mean core radius and polydispersity $152 \pm 2$~nm and $6.1 \pm 2\%$ respectively, and volume fraction 0.006.

The fitted mean thickness of the PHSA hair was $6 \pm 1$~nm, agreeing with the value obtained by Cebula et al. from core-shell fitting of small-angle neutron scattering data \cite{Ottewill83} and viscosity measurements \cite{Barsted71}. Our core-shell fitting (and that in reference \cite{Ottewill83}) assumed a homogeneous shell, while viscosity measurements return a value relevant for hydrodynamic properties. Cebula et al. also measured structure factors of concentrated suspensions, and from the fitted effective hard-sphere diameter deduced a shell thickness of 10~nm. They attributed the difference with core-shell fitting the solvation of the shell and the polydispersity of the PHSA chains. Indeed, chains as long as 15-20~nm may be present \cite{Barsted71}. The longer chains will have a significant effect in determining an effective hard-sphere diameter, which presumably is the relevant characterization for our purposes. We therefore take the shell thickness to be $10$, with a polydispersity of $\pm 4$~nm. This returns a final estimated particle size of $162$~nm and a polydispersity of 6.2\% (the latter figure having been obtained by adding the variances of the core and shell).

The polymer was linear polystyrene (PS) with mass-averaged molecular weight $M_{\rm W} = 114,200$~daltons (Aldrich GPC Standard, unknown polydispersity), with an estimated radius of gyration of $r_g \approx 10$~nm at room temperature, where we expect PS to behave as nearly-ideal coils in our solvent \cite{Berry66}.  Thus the (average) size ratio of our system is $\xi \approx 10/158 = 0.062$. We estimate that the overlap concentration, $c^*$, is reached at a polymer concentration of $c_p \approx 45$~mg/cm$^3$. The highest concentration we used was $\lesssim 0.4c^*$. 

\begin{figure}[t]
\begin{center}
\includegraphics[width=0.7\textwidth,clip]{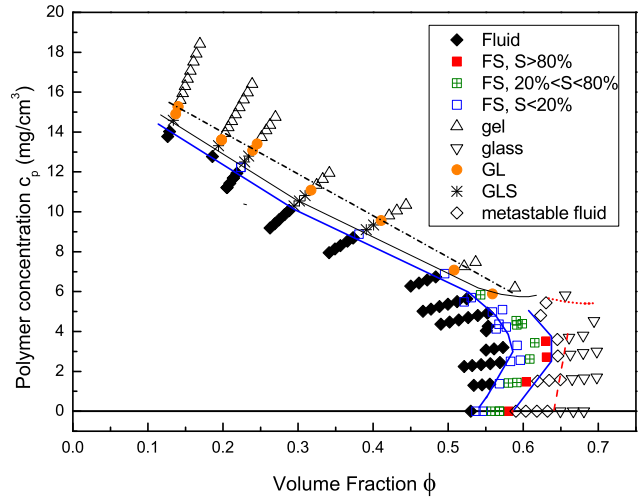}
\end{center}
\caption{Observed phase diagram, with legend given in the inset, where G, L and S stand for gas, liquid and solid (= crystal) phases. Three kinds of FS coexistence samples are distinguished, those with $< 20\%$, $> 80\%$ or between 20-80\% of crystals: S~$< 20\%$, S~$ > 80\%$ and $20\% < {\rm S} < 80\%$. Lines are guides to the eye of where various phase boundaries lie as deduced from the data points; the exception is the dotted line, which is a schematic of the attractive glass line. (Colour online.)}
\label{phdiag}
\end{figure}

We calibrated the colloid volume fraction using the theory of Fasolo and Sollich for the phase behaviour of polydisperse hard spheres \cite{Sollich03}. The latter predicts that in the vicinity of $\sigma \approx 6\%$, the solid branch of the FS coexistence region displays a maximum at $\phi = 0.58$. We assigned $\phi = 0.58$ to a pure colloid sample with close to 100\% crystals (\Red{$\blacksquare$} on the $c_p = 0$ axis, Fig.~\ref{phdiag}). This procedure introduces systematic uncertainties into our $\phi$ values in the region of $1\%$ or more, although the statistical uncertainties in the quoted $\phi$ values are considerably smaller.

Mixtures were prepared in 1.8~ml cuvettes by adding solid polymer to colloidal suspensions with known $\phi$ and tumbling overnight. Phase behaviour was studied by direct visual observation over 3 to 14 days at 25$^{\circ}$C. Time-lapsed images were also taken in a number of cases. After observation, samples were homogenized again by tumbling, and then diluted to yield the next generation of samples for observation. All our data points therefore lie along `dilution lines' in the $(\phi, c_p)$ plane. For convenience, however, we describe our observations below in terms of the progressive addition of polymer to samples. 

\section{Results}

The different kinds of behaviour we observed are summarised in a phase diagram in Fig.~\ref{phdiag}. At $\phi \lesssim 0.55$ and low enough polymer concentration, $c_p$, samples remained single-phase fluids ($\blacklozenge$ in Fig.~\ref{phdiag}). Adding sufficient polymer brought about fluid-solid (FS) coexistence (\Blue{$\square$} in Fig.~\ref{phdiag}), with iridescent crystallites nucleating throughout a sample and then falling to the bottom of the cuvette. Further addition of polymer gave rise to samples that separated into two coexisting, non-crystalline phases with different turbidities (and therefore volume fractions) in which the interface between the phases was sharp and remained horizontal when a sample was tilted (\Orange{$\bullet$} and $\ast$ in Fig.~\ref{phdiag}). This is the behaviour expected of samples undergoing gas-liquid (GL) phase separation. A number of samples initially showing GL phase separation subsequently nucleated crystals from the denser, liquid phase, giving rise to final states displaying gas, liquid and crystal (GLS) phases ($\ast$ in Fig.~\ref{phdiag}). 

At even higher polymer concentrations in this region, the behaviour was observed to change again ($\triangle$ in Fig.~\ref{phdiag}). The final state of these samples looked at first sight very similar to those that had undergone GL phase separation. But the interface between upper and lower phases was often visibly rough, and did not tilt when a sample was tilted. In some cases, we obtained sedimentation profile from time-lapse images; these showed a delay time before rapid sedimentation began. Such `delayed sedimentation', giving rise to a solid-like sediment, is characteristic of transient gelation \cite{Poon95b}.

In the region $0.55 \lesssim \phi \lesssim 0.58$ and $c_p = 0$, we again found FS coexistence. The amount of crystals grew as $\phi$ increased (\Green{$\boxplus$}), until we reached a sample that was essentially fully crystalline (\Red{$\blacksquare$} in Fig.~\ref{phdiag}), which delimits the high-concentration side of the FS coexistence region. Upon addition of polymer, the width of the FS coexistence region stayed roughly constant, but displayed a re-entrant shape. Interestingly, all samples to the immediate right hand side of this region failed to crystallise ($\lozenge$ and $\triangledown$ in Fig.~\ref{phdiag}). This is in striking contrast to what was observed in a colloid-polymer mixture containing colloids with a somewhat lower $\sigma$ of 5\% \cite{Poon93}, where immediately to the right of the FS coexistence region, a region of full crystallisation was found. On the other hand, our observations are consistent with previous work on another mixture with a $\xi$ similar to the present one and $\sigma \approx 7\%$ \cite{Poon02b,Poon04b}. In that work, it was assumed that all non-crystallizing samples to the right of the FS coexistence region were glassy \cite{Poon02b}, although it was noted that this assumption might not be correct \cite{Poon04b}. An unambiguous method for deciding whether a non-crystallizing sample is a fluid or a glass is dynamic light scattering, which measures the intermediate scattering function (ISF). The ISF decays completely in fluids, but has a finite value in the long-time limit in glasses. But our samples are turbid, and would require the use of specialised light scattering equipment to probe their ISF \cite{Pusey99}. We therefore chose a less precise, but much easier to implement, visual method. Samples whose menisci stayed horizontal when vials were tilted were classified as fluids, while samples whose menisci tilted immediately with the vials and stayed tilted for some time were classified as glasses.

Since we classified samples by visual observations alone, the boundaries between the different types of behaviour described above are not entirely sharp. Nevertheless, taking into account uncertainties in observation and in sample concentration, we can locate these boundaries to within about one data point in a typical dilution sequence. The approximate position of the boundaries demarcating the different types of behaviour are sketched in Fig.~\ref{phdiag}.

\section{Discussion}\label{Discuss}

\begin{figure}[t]
\begin{center}
\includegraphics[width=0.32\textwidth,clip]{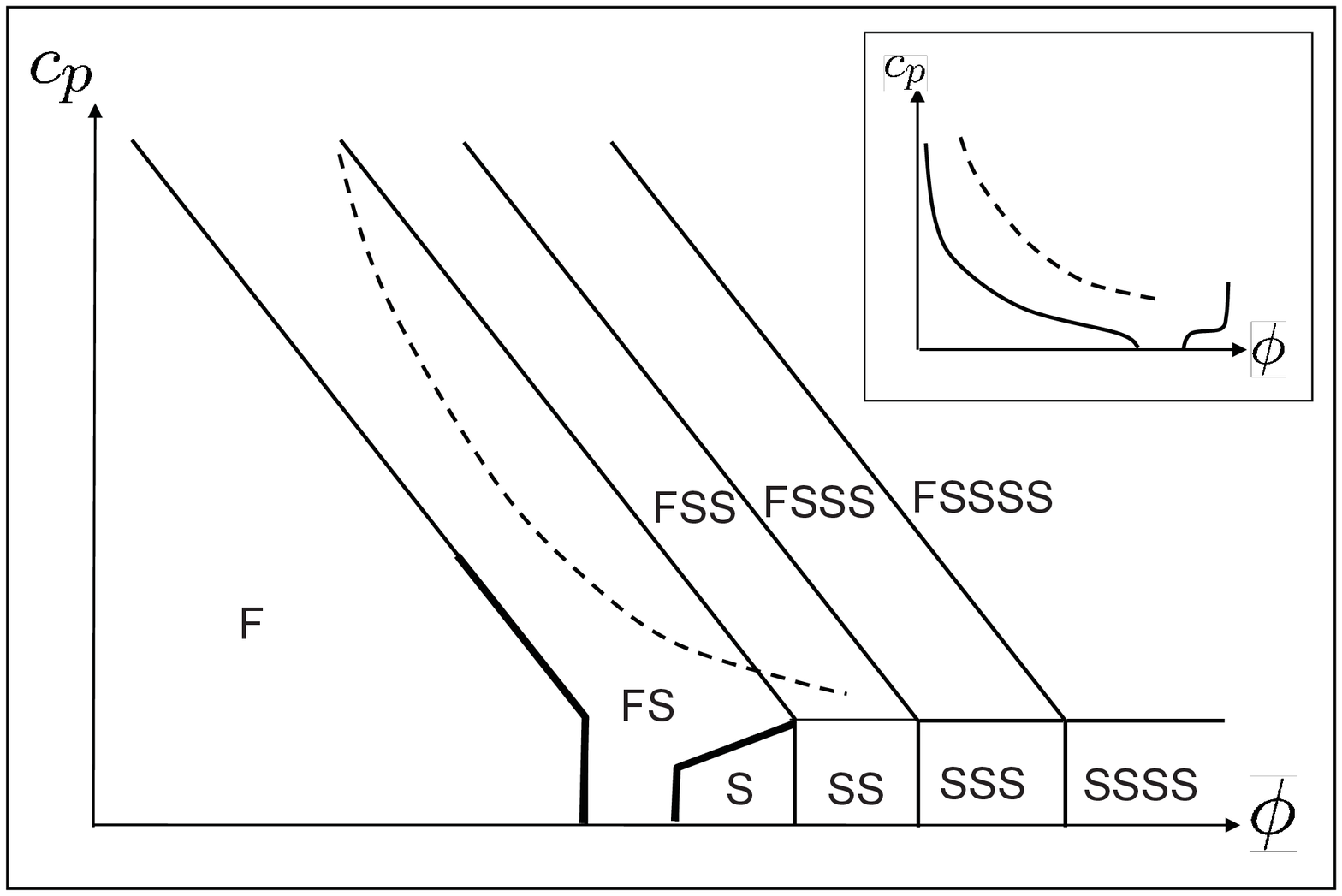}
\includegraphics[width=0.32\textwidth,clip]{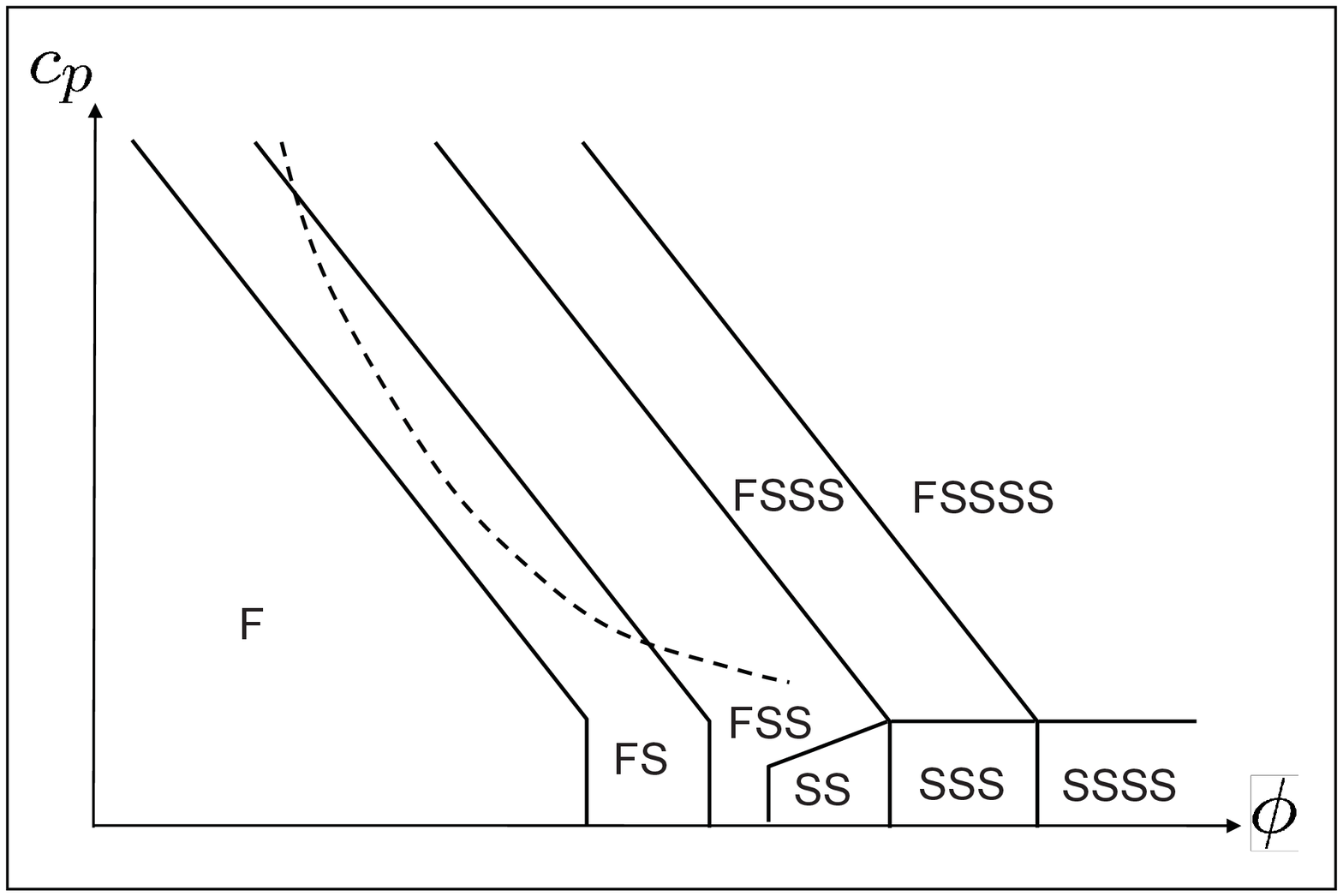}
\includegraphics[width=0.32\textwidth,clip]{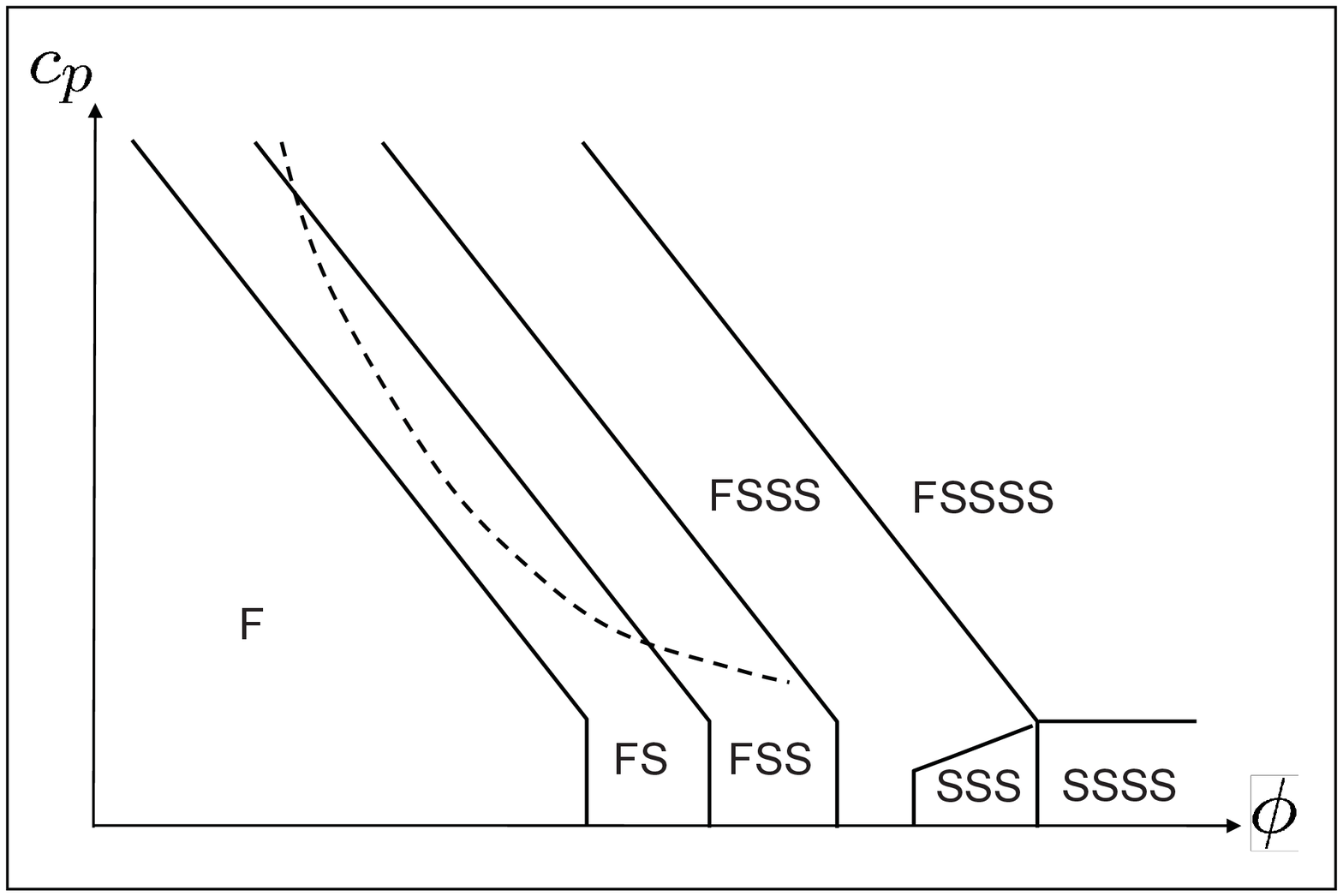}
\end{center}
\caption{Topology of the phase diagram of a colloid-polymer mixture with $\xi$ appropriate to our experimental system according to \cite{Sollich05}. (a) $\sigma \lesssim 0.6$. The bold parts show what is observed experimentally. Inset: a schematic representation of the phase diagram for a mixture of non-adsorbing polymers and monodisperse colloids. (b) $0.6 \lesssim \sigma \lesssim 0.75$. (c) $\sigma \gtrsim 0.75$. In all cases, we have added a metastable gas-liquid binodal (dashed curve).}
\label{sollich}
\end{figure}

We interpret our data within the framework of the possible phase diagram topologies in polydisperse colloid + polymer mixtures given by Fasolo and Sollich \cite{Sollich05}. For low enough $\xi$, and certainly for $\xi < 0.2$ (compare Figs.~2 and 5 in \cite{Sollich05}) the topologies for low, medium and high polydispersities are shown in Fig.~\ref{sollich} (redrawn from Fig.~5 in \cite{Sollich05}). From the sequence of phases expected for pure hard sphere (i.e. $c_p = 0$) as $\sigma$ increases (see in particular Fig.~3 in \cite{Sollich05}), we may deduce that Fig.~\ref{sollich}(a) should apply for $0 < \sigma < \sigma_{c1} \approx 6^{+}\%$, Fig.~\ref{sollich}(b) for $\sigma_{c1} < \sigma < \sigma_{c2} \approx 8^{-}\%$, and Fig.~\ref{sollich}(c) for $\sigma > \sigma_{c2}$. In other words, Fasolo and Sollich's theory predicts sharp transitions between different phase diagram topologies at well defined values of the polydispersity, $\sigma_{c1}$, $\sigma_{c2}$, etc. In the vicinity of these `critical polydispersities', a small change in $\sigma$ can give rise to qualitative changes in the observed phase diagram.

Note now from Fig.~\ref{sollich}(a) that when the polydispersity is low enough, $\sigma < \sigma_{c1}$, the low $c_p$ portion of the phase diagram, picked out by bold lines in the figure, is topologically identical to that predicted for a mixture in which the colloids are monodisperse (inset, Fig.~\ref{sollich}(a)). Previous work on colloid-polymer mixtures at low $\xi$ and low $\sigma$ \cite{Poon95,Poon93} confirmed this prediction. Addition of polymer was found to broaden the FS coexistence region. At higher polymer concentration, various non-equilibrium phenomena were encountered, which could be interpreted as due to the presence of a metastable gas-liquid (GL) binodal (dashed curve, Fig.~\ref{sollich}(a) and inset) `buried' within the equilibrium FS coexistence region. 

No such broadening of the FS coexistence region is seen in our phase diagram, Fig.~\ref{phdiag}. Instead, the FS coexistence region remains more or less constant in width, although it has an interesting re-entrant shape. The behavior is consistent with the topology shown in either Fig.~\ref{sollich}(b) or (c). Since our measured polydispersity is just over 6\%, it is probable that Fig.~\ref{sollich}(b) is applicable. 

But whether Fig.~\ref{sollich}(b) or (c) applies in our case is in fact irrelevant for our purposes. In either case, the predicted equilibrium phase behaviour to the high-$\phi$ side of the FS coexistence region is {\em not} single-phase solid, but the coexistence of multiple solid phases with fluid. In recent simulations of polydisperse hard spheres \cite{Pusey09}, it was found that the locus of points beyond which crystallization failed to occur closely tracked Fasolo and Sollich's theoretical boundary \cite{Sollich03} separating the occurrence of a single solid (S) and two solid coexistence (SS) in the $(\phi, \sigma)$ plane. Apparently, fractionation into multiple solid phases is kinetically unattainable (cf. \cite{vanMegen03}). Our observations suggest that the same may be true in colloid-polymer mixtures: all samples immediately to the high-$\phi$ side of our FS coexistence region remained in the fluid state rather than show FSS coexistence. 

At even higher volume fractions, samples became glassy. The boundary separating fluid from glass appear to have a slight positive slope, suggesting re-entrance, i.e. the addition of polymer (= interparticle attraction) to the hard-sphere glass initially melts it into a metastable fluid, as found before \cite{Poon02b,Poon04b}. 

At $\phi \lesssim 0.5$ we find a FS cloud curve (i.e. the low-$\phi$ limit of the FS coexistence region) with negative slope, consistent with any of the topologies in Fig.~\ref{sollich} and with previous experiments at similar \cite{Poon02b} or lower \cite{Poon93} polydispersities. Inside the FS coexistence region, we find evidence for a metastable GL binodal, again consistent with previous work \cite{Poon95,Poon02b}. Such a binodal is known to be `buried' within the FS coexistence region in the monodisperse system (inset, Fig.~\ref{sollich}(a)), becoming stable in its own right at $\xi > \xi_c$ \cite{Poon92}. 
A metastable GL binodal clearly continues to exist in polydisperse systems with low $\xi$.
Inside such a metastable binodal (dotted lines in Fig.~\ref{sollich}), a homogeneous system may lower its free energy somewhat by separating into coexisting gas and liquid phases, although the absolute minimum in free energy (i.e. thermodynamic equilibrium) is only reached by separating into one or more of the solid phases indicated by the equilibrium phase boundaries (full lines in Fig.~\ref{sollich}). (For further discussion of `metastable phase boundaries' of this kind, see \cite{deHoff}.)

 Compared to a system with lower $\sigma$ \cite{Poon93}, the region of FS coexistence between the FS cloud curve and the metastable GL binodal in our higher-$\sigma$ system is very narrow (compare also \cite{Poon02b}). This is consistent with the finding that polydispersity destabilises the single-phase fluid relative to GL phase separation \cite{Wilding04}, so that less polymer is needed to cause phase separation \cite{Sollich05}. 

It is by now established that arrested phase separation within a GL binodal gives rise to gelation \cite{Zaccarelli08}. In colloid-polymer mixture with $\xi < \xi_c$, such arrest is due to the intersection of the attractive glass line \cite{Poon02b} with the GL binodal \cite{Zaccarelli08}, so that the gelation boundary is a GL tie line. In systems with low enough $\sigma$, this intersection occurs very close to the GL critical point \cite{Zaccarelli08}, so that a region of complete GL phase separation is not observed \cite{Poon93,Zaccarelli08}. However, with the movement of the GL critical point to lower $c_p$ in a system with higher $\sigma$, a region of complete (i.e. non-arrested) GL phase separation can be seen, as is the case in our system here. Significantly, this region is separated from the region of transient gelation (= arrested GL phase separation) by a straight line, Fig.~\ref{phdiag}: this is a tie line within the GL binodal. The high-$\phi$ end of this tie line indeed lies where we may expect the attractive glass line (shown schematically as the dotted line in Fig.~\ref{phdiag}) to intersect the GL binodal . 

A number of samples just across the metastable GL binodal displayed interesting multi-stepped kinetics ($\ast$ in Fig.~\ref{phdiag}): they first separated into coexisting gas and liquid phases; subsequently, crystals nucleated out of the liquid phase, giving rise to samples displaying gas, liquid and solid phases. We interpret this as a non-equilibrium fractionation effect. The liquid phase is expected to have lower polydispersity than the parent metastable fluid \cite{Sollich05,Evans98,Wilding04}. This lower polydispersity is favourable for crystal nucleation. However, equilibrium FS coexistence cannot be attained, because a non-equilibrium particle size distribution was `locked into' the liquid phase, and then into the crystals that nucleated from this phase. The (experimental) final state of these samples therefore contains out-of-equilibrium gas, liquid and crystal phases. These samples must be carefully distinguished from those showing equilibrium gas, liquid, solid coexistence in colloid-polymer mixtures with $\xi > \xi_c$ \cite{Poon95}.

However, this two-stepped crystallization mechanism is only observed for samples close to the GL binodal. Here, GL phase separation is at its slowest, so that a higher level of fractionation should be kinetically attainable; this is favourable to crystallization from the liquid phase. Deeper into the GL binodal, GL phase separation occurs faster, so that we may expect a lower degree of fractionation on kinetic grounds. We suggest that this kinetic effect eventually renders the liquid phase not sufficiently fractionated to give rise to crystallization. Metastable GL coexistence is therefore the (experimental) final state of these samples. 

\section{Summary and outlook}

We have studied experimentally the phase behaviour of a mixture of hard-sphere-like colloids and non-adsorbing polymers in which the colloids have a polydisersity of $\sigma \gtrsim 6\%$ and the polymer:colloid size ratio is $\xi \approx 0.062$. The observed phase diagram displays a number of features not found in similar mixtures with lower colloid polydispersity.\footnote{Note that the refractive index mismatch between particles and solvent here is closely similar to that in \cite{Poon95,Poon93}, so that differences are not due to differing residual van der Waals interactions.} Most notably, the FS coexistence region does {\em not} expand upon the addition of polymer, but rather takes a striking re-entrant shape of constant width. The physical origins of this re-entrant FS coexistence region is unclear at present. 

Inside a metastable GL binodal, we observed two-stepped crystallization kinetics: samples first separated into coexisting gas and liquid phases, and crystals subsequently nucleated from the liquid phase. We have interpreted this as a kinetic fractionation effect: a sufficiently fractionated liquid phase can nucleate crystals. Experiments in which the degree of fractionation in coexisting phases is measured, e.g. by fitting form factors from SAXS, can be used to confirm or refute this interpretation. Note that the multi-stepped kinetics we observed is consistent with a phenomenological model used previously to interpret observations from a colloid-polymer mixture with $\xi > \xi_c$ so that equilibrium GLS coexistence was possible \cite{Renth99,Renth01}. Moreover, if our interpretation of these observations in terms of fractionation is correct, then our results can be seen as a novel application of the Ostwald `Rule of Stages' \cite{Ostwald} in polydisperse systems.

The constant-width FS coexistence region we observed is consistent with predictions of the phase diagram topology by Fasolo and Sollilch \cite{Sollich05} for colloid-polymer mixtures with our $\xi$ and $\sigma$. On the other hand, the regions of multiple solid coexistence predicted by this equilibrium theory are replaced in our experiments by metastable fluids or glasses --- fractionated solid-solid coexistence is apparently kinetically inaccessible, either in mixtures of hard spheres and polymers (this work) or in pure hard spheres \cite{Pusey09}. Such kinetic effects have not been predicted by any first-principle theory to date. Note that interestingly, coexisting hexagonal columnar crystal phases {\em have} recently been observed in mixtures of polydisperse colloidal platelets \cite{Byelov10}, although only after very long aging times (2 to 5 years). It is therefore possible that future experiments in hard spheres or hard spheres + polymer mixtures over similarly long time scales (years) will observe FSS or SS coexistence experimentally.

Our intention here has not been to study the effects of polydispersity on the re-entrant glass transition known to occur in colloid-polymer mixtures with low $\xi$ and $\sigma$ \cite{Poon02b,Poon04b}. Indeed, such effects have not yet been probed in any detail, although one experimental report in a system with $\sigma \approx 23\%$ indicates that such a high polydispersity may obliterate the re-entrance all together \cite{Poon08b}. Our observations suggest that the re-entrance is still present in a system with $\sigma \gtrsim 6\%$, although more data points and a detailed study using dynamic light scattering will be necessary to confirm this finding. It would also be interesting to probe the dynamics of the metastable fluids ($\lozenge$ in Fig.~\ref{phdiag}) that seem to be prevented by kinetics from reaching equilibrium states in which multiple solid phases should coexist \cite{Sollich05}.

\subsubsection*{Acknowledgements} WCKP was funded by the EPSRC (EP/D071070/1), and SML holds an EPSRC studentship. We thank Mike Cates and Peter Sollich for helpful discussions, and Paul Clegg for suggesting the Edinburgh-ESRF collaboration.

\end{document}